# Reputation-Based Federated Learning Defense to Mitigate Threats in EEG Signal Classification

Zhibo Zhang
C2PS, Department of Electrical Engineering and Computer Science
Khalifa University
Abu Dhabi, United Arab Emirates
100060990@ku.ac.ae

Pengfei Li
Department of Electrical Engineering and Computer Science
Khalifa University
Abu Dhabi, United Arab Emirates
100062567@ku.ac.ae

Ahmed Y. Al Hammadi
C2PS, Department of Electrical Engineering and Computer Science
Khalifa University
Abu Dhabi, United Arab Emirates
ahmed.yalhammadi@ku.ac.ae

Fusen Guo
School of Information Technology
Deakin University
Geelong, Australia
dobbysen430@gmail.com

Ernesto Damiani
C2PS, Department of Electrical Engineering and Computer Science
Khalifa University
Abu Dhabi, United Arab Emirates
ernesto.damiani@ku.ac.ae

Chan Yeob Yeun
C2PS, Department of Electrical Engineering and Computer Science
Khalifa University
Abu Dhabi, United Arab Emirates
chan.yeun@ku.ac.ae

*Abstract*—This paper presents a reputation-based threat mitigation framework that defends potential security threats in electroencephalogram (EEG) signal classification during model aggregation of Federated Learning. While EEG signal analysis has attracted attention because of the emergence of brain-computer interface (BCI) technology, it is difficult to create efficient learning models for EEG analysis because of the distributed nature of EEG data and related privacy and security concerns. To address these challenges, the proposed defending framework leverages the Federated Learning paradigm to preserve privacy by collaborative model training with localized data from dispersed sources and introduces a reputation-based mechanism to mitigate the influence of data poisoning attacks and identify compromised participants. To assess the efficiency of the proposed reputation-based federated learning defense framework, data poisoning attacks based on the risk level of training data derived by Explainable Artificial Intelligence (XAI) techniques are conducted on both publicly available EEG signal datasets and the self-established EEG signal dataset. Experimental results on the poisoned datasets show that the proposed defense methodology performs well in EEG signal classification while reducing the risks associated with security threats.

## I. INTRODUCTION

The rapid advancement of brain-computer interface (BCI) technology has led to an increasing interest in collecting and examining electroencephalogram (EEG) data [1]. Among the applications of the EEG signal data, utilizing EEG signals to identify human emotional states is an effective way to detect possible internal threats posed by insiders, such as current or former employees [2]. Moreover, the ongoing advancements in artificial intelligence (AI) technologies, such as Deep Learning and machine learning, have contributed to the recent improvements in classifying human emotions [3]. However, developing effective learning models for EEG analysis faces significant challenges due to the distributed nature of EEG data, along with privacy concerns and security threats [4]. Therefore, the aim of this paper is to propose a reputation-based federated learning framework, specifically tailored to overcome these challenges and improve the security and privacy of EEG signal analysis.

The motivations behind this research work are to provide a reliable and efficient mechanism to deal with the distributed and security-preserving nature of EEG data. EEG data is usually collected from multiple sensors or electrodes located on different parts of the scalp or even from multiple individuals [5]. This leads to a decentralized nature of EEG data, which can be difficult to manage and process [6]. On the other hand, privacy [7] is another significant concern when it comes to the EEG data collection and storage due to the potential of revealing sensitive information related to an individual's health, mental state, thoughts, and emotions. The misuse or unauthorized access to EEG data can result in negative consequences such as discrimination or stigmatization [8]. Besides, the security threat is another issue that should be taken into consideration. Since EEG data is usually stored and transmitted digitally, it is susceptible to hacking and other cyber threats [9]. In some applications where EEG data is utilized to operate devices or communicate with computer systems, attackers may exploit EEG signals to hijack the system or access sensitive information [2], [10].

Federated Learning was proposed by McMahan *et al.* [11] [12] from Google to deal with the security and decentralization issue in the application of machine learning. Federated learning has gained significant attention from academia and industry as it allows for the training of machine learning models without centrally stored data. Instead, participants train models locally with their own data and send only the local models without raw data to the model aggregation server. This approach allows for the creation of an accurate global model while keeping personal data secure. Although machine learning has been widely used in the analysis of EEG data, the rise of Federated Learning has the potential to address the existing challenges related to privacy, decentralization, and security in the field of EEG signal classification [13], [14].

On the other hand, Federated Learning is vulnerable to security risks, especially from data poisoning attacks that can harm the accuracy of the trained model as the global node has no information about the training process of the local nodes [15], [16]. Furthermore, EEG data is known to be unreliable, and there is a possibility of unintentional actions by data owners, both of which can result in deviations in certain updates [17]. To tackle these problems, this paper introduced a solution utilizing a reputation-based approach to handle threats and ensure the integrity of the model during the aggregation process, specifically developed for EEG signal classification in Federated Learning.

As motivated, this paper aims to provide a robust and effective reputation-based data poisoning defending framework. To this goal, this work first developed a training data poisoning attacker based on the data risk level utilizing label flipping and feature manipulation. The data risk level and feature importance level of the training EEG data were annotated by the Support Vector Machine (SVM) and Explainable Artificial Intelligence (XAI) techniques respectively. Then, the proposed defending Federated Learning framework employing the reputation-based mechanism on the reputation of local EEG data nodes was conducted to fight against the previous SVM-XAI attacker. The contributions of this paper are summarized as follows.

1) This paper develops a training data poisoning mechanism based on label flipping and data risk level named SVM-XAI attacker. Under this mechanism, the performance of EEG data classification based on machine learning degrade.
2) Considering the security and privacy requirements of the EEG data classification tasks, a Federated Learning framework rather than conventional machine learning classifiers is established for emotion classification.
3) Reputation rules were deployed in the model aggregation process of Federated Learning to evaluate the security of local nodes. The vulnerabilities of the Federated Learning training process are mitigated using reputation rules.
4) Based on the experiments on both the self-established EEG signal dataset and the state-of-the-art datasets, the proposed reputation-based Federated Learning framework can defend and mitigate the data poisoning threats in EEG signal classification.

The remainder of this paper is organized as follows. The related work is discussed in Section 2. The SVM-XAI training data poisoning attacker is introduced in Section 3. The reputation-based federated learning defending algorithm is presented in Section 4. Experimental simulation results are shown in Section 5. The conclusion and future work of this paper are in Section 6.

## II. RELATED WORK

In this section, the existing related research papers on the topic of federated learning security and EEG signal data classification are reviewed.

Some research works utilizing traditional machine learning methods to evaluate EEG signal data are reviewed. Ahmad et al. in [18] suggest a machine-learning framework that utilizes electroencephalogram (EEG) signals to objectively evaluate human stress levels. The proposed method could measure stress levels accurately and could aid in developing a computer-assisted diagnostic tool for stress detection. The study uses the fast Fourier Transform (FFT) to estimate the EEG absolute power, while the relative power is obtained by deriving it from the absolute power of the frequency bands. Considering emotion recognition using EEG data, Huang et al. in [19] an EEG-based BCI system for emotion recognition training using the instant feedback EEG data induced from positive and negative video clips. Research papers are working on implementing deep learning methods in EEG signal data processing recently as well. In [20], Ahmed et al. evaluate fitness based on the mental state of an individual and classify it into four categories using a risk matrix through a deep learning algorithm. Eldele et al. in [21] present the AttnSleep, a new deep learning model for sleep stage classification using single channel EEG signals. The proposed model incorporates an attention mechanism and utilizes a multi-resolution convolutional neural network (MRCNN) and adaptive feature recalibration (AFR) for feature extraction.

In addition to the literature on security concerns of EEG signal data, other research on defending the EEG signal analysis framework is also emerging. Zhang et al. in [2] exploit that data poisoning attacks on systems used for evaluating human emotions based on EEG signals could be explained using various Explainable Artificial Intelligence (XAI) methods such as Shapley Additive Explanation (SHAP) values and Local Interpretable Model-agnostic Explanations (LIME). In [6], the authors utilize long short-term memory (LSTM) to analyze brainwave signals and remember the past mental states of an individual. The presented system then compares these past states with the current state of the brain to classify the associated risk level to protect the safety of critical industrial infrastructures. In [22], Xiao et al. develop two sets of innovative PoC attacks that can be easily utilized, comprising four remote attacks and one proximate attack to infer users' activities based on the reduced-featured EEG data stolen from IoT devices.

While the previous works highlight the importance of safety and privacy issues in EEG signal data processing, some marvelous works introduce federated learning in this area. Xu et al. in [4] expand the application of Federated Learning to the domain of EEG signal-based emotion recognition and assess its precision on the SEED and DEAP datasets reaching a model accuracy of 90.74%. This work also demonstrates the necessity

The Authors gratefully appreciate the support from the Khalifa University and Technology Innovation Institute (TII) under Grant PALM 8434000394.

of Federated Learning methods in emotion classification by comparing the performance of clients training with only local data. In [7], Ce Ju et al. suggest a new privacy-focused deep learning architecture called Federated Transfer Learning (FTL) for EEG classification, which is built upon the federated learning framework. The proposed architecture uses domain adaptation techniques to extract common discriminative information from multi-subject EEG data, working with the single-trial covariance matrix. In terms of privacy-preserving, Mohd et al. in [13] utilize an Artificial Neural Network (ANN) as a baseline model for the classification of emotional states, specifically Arousal, Valence, and Dominance. The proposed framework, FedEmo, incorporates the concept of federated learning to preserve privacy, allowing for local training on the clients' end while keeping the model updated through a global server without any privacy breaches.

On the other hand, the vulnerabilities of the Federated Learning method itself are also highlighted by research works. Bouacida et al. in [15] aim to fill a gap in the literature on Federated Learning by conducting a thorough survey of the various security vulnerabilities that are inherent in the Federated Learning ecosystem. In [23], Qi et al. proposed a blockchain-based federated learning (BFL) framework to mitigate the influence of malicious nodes. The process of federated learning can be conducted fairly and transparently through smart contracts on the blockchain. Additionally, a mechanism has been designed to encourage data owners to participate in BFL by rewarding them for contributing high-quality data. This mechanism involves limiting the contribution of data based on the reputation of the data owner. Sun et al. [24] introduce a new optimization method called "attack on federated learning" (AT2FL) that takes into account the structure of the system being attacked. This method allows for the efficient derivation of implicit gradients for poisoned data, which can then be used to find optimal attack strategies in federated machine learning.

All the above research works are marvelous solutions. They implement state-of-the-art artificial intelligence techniques including machine learning, deep learning, and federated learning in the tasks of EEG signal data classification. Considering the privacy and safety issues of EEG signal data, some research works play as attackers whereas some papers defend against the attackers to investigate the vulnerabilities of the frameworks of EEG signal analysis. However, most of the existing solutions do not consider the efficiency of EEG data attackers and the local nodes' safety in federated learning. Therefore, based on the investigations of the existing works, this paper takes the research a step further. More specifically, introduces a method for training data poisoning referred to as SVM-XAI attacker, which involves label flipping and data risk level. This method results in a degradation of the performance of EEG data classification based on machine learning. Furthermore, this work introduces a reputation-based mechanism for the defense of malicious data poisoning in federated learning local nodes. Whereas suspicious nodes will be removed from the global model, the influence of nodes with less reputation on the global model will also be reduced.

TABLE I. MAJOR NOTATIONS

| Notation | Explanation |
|---|---|
| $D_k, D'_k, D^*_k$ | Original EEG data set in node k, EEG data set with risk level assessment annotations in node k, corrupted EEG data set in node k |
| $L$ | Linear separator |
| $R$ | Risk level set |
| $x_i, y_i$ | A piece of EEG data and its label in the data set |
| $sv_j$ | Support vectors |
| $E$ | XAI explainer, feature permutation importance |
| $\alpha$ | Attacking budget |
| $flag_k$ | Attacking flag |
| $f_{max}, f_{min}$ | The most and least important features |
| $r_k$ | Initial reputation in node k |
| $\eta$ | Learning rate |
| $w^0, w^T$ | Initial global model parameter and final global model parameter |
| $w_k^0$ | The initial local parameter in node k |
| $T, G$ | Aggregation round and group |
| $e_k^{t+1}$ | Contribution value of node k in aggregation round t+1 |
| $e_{min}$ | Threshold of the contribution value |

III. SVM-XAI ATTACKER

This section presents a training data poisoning attacker based on label flipping and feature manipulation. Label flipping attack is accomplished by risk level evaluation using SVM whereas feature manipulation attack is achieved by the XAI technique, feature permutation importance. The most significant EEG signal data and electrode features would be evaluated by SVM-XAI and attacked by label flipping and feature modification. For convenience, the major notations are concluded in Table I.

*A. SVM-based Risk Level Assessment*

Estimating the gravity and likelihood of an attack on an asset is necessary for the risk assessment [25]. The likelihood is based on known threats and vulnerabilities, whereas the severity is based on the value of the affected assets. To estimate the risk, the product of severity and likelihood is frequently utilized [26]. The idea of a risk index, which quantifies severity and likelihood into discrete levels or a continuous score, has just been suggested. To calculate the risk score of each unique data point in the case of data tampering risk, machine learning models can be used to learn a function on the data space [27]. Numerous techniques rely on the linear regression model, in which the risk score changes linearly with the size of the data set [28]. However, because close data points may differ in attack severity or likelihood, linear approaches may not be appropriate for many types of data [29].

In this work, shown in the pseudo-code of Algorithm 1, the risk level of each point in the training dataset is calculated by

**Algorithm 1** Risk Level Assessment Algorithm

**Input:** Original training set in each node $D_k = \{(x_i, y_i)\}_{i=1}^n (k = 1,2 \dots, K)$, linear separator L, risk level set R.

1: $j \leftarrow 1$
2: **for** k = 1 to K **do**
3:    num = $|D_k|$
4:    **while** j < num **do**
5:      Define support vectors $sv_j$ of $L_j$ separating data points in $D_k$
6:      Distribute risk level $r_j$ to $sv_j$
7:      Add $\{sv_j, r_j\}$ to $D_k$
8:      Delete $sv_j$ from $D_k$
9:      $j \leftarrow j + 1$
10:   **end while**
11: **end for**
12: $D'_k \leftarrow D'_k$

**Output:** Training set in each node with risk level in each point $D'_k = \{(x_i, y_i, r_i)\}_{i=1}^n (k = 1,2 \dots, K)$.

determining whether or not it belongs to the set of support vectors of an SVM that was trained on the entire dataset. Points that are closer to the SVM separating hyperplanes [30] are considered more important. The process we use to identify support vectors involves iteration shown in Algorithm 1, where we progressively prune the points that support the separator in a previous iteration until we can identify the support vectors of the hyperplanes that separate classes. Fig. 1 shows the method of assessing varying levels of risk based on how close data points are to separator hyperplanes in the proposed algorithm.

### B. Attack Model

The assumptions about the attacker's abilities and their familiarity with the targeted machine learning models are outlined, followed by a comprehensive explanation of the strategy and the form of poisoning attack that they can execute in this section. The attacker's goal would be to manipulate the most vulnerable data points. However, their knowledge of the targeted machine learning models is limited. Under this circumstance, rational attackers would take the strategies shown in algorithm 2 to choose the samples and features that would result in the greatest reduction in accuracy of the targeted model if altered.

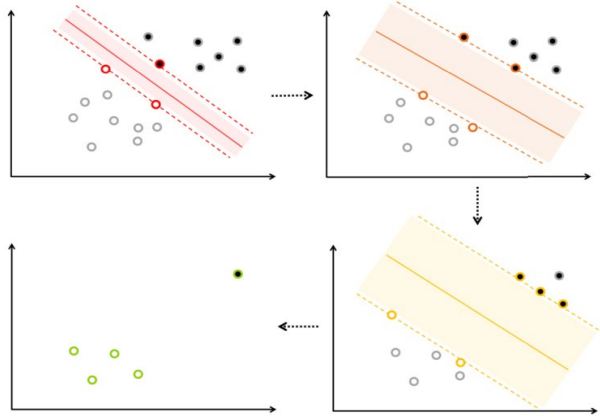

Fig. 1. A workflow of how to evaluate the risk level of data points using separator hyperplanes.

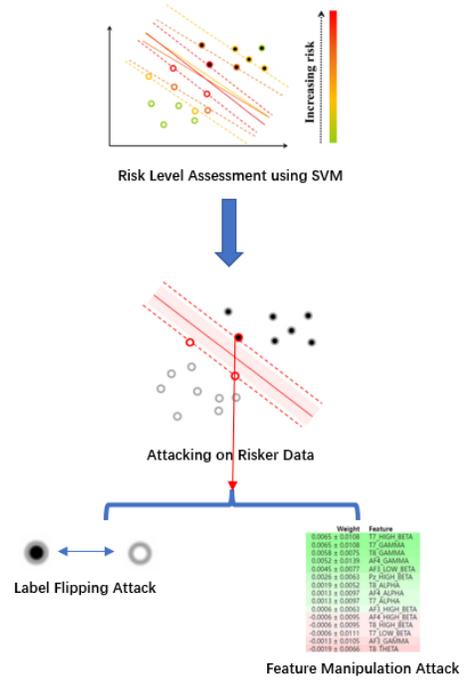

Fig. 2. A overview of a rational attacker's strategy utilized in Algorithm 2.

The overview of the strategies that a rational attacker would take on the EEG training data according to Algorithm 2 is shown in Fig. 2. Although the attackers could launch the label flipping attack based on the risk level assessment results according to algorithm 1, it is still confusing for attackers to conduct the feature modification attack as the feature importance is not revealed. Therefore, this work introduces the XAI technique to identify the feature importance level in the collected EEG signal data.

Due to the "black box" manner of most machine learning and deep learning methods, incorporating XAI is becoming crucial when developing cyber security models to ensure the models are both accurate and comprehensible to human users in recent years [31]. This will enable users to trust and manage the next generation of cyber defense mechanisms. On the other hand, XAI methods could be implemented from the attackers' side to the vulnerabilities of machine learning [2] and deep learning [32] cyber defense systems as well. Therefore, in this work, XAI methods, feature permutation importance would be implemented by rational attackers to evaluate feature importance and operate the feature modification attack. The XAI technique of feature permutation importance evaluates the significance of each feature in a model by permuting its values in the test set and measuring the decrease in the model's accuracy. This method identifies the most important features for the model's prediction, making it particularly useful in EEG signal

**Algorithm 2** Attacker's Strategy Using Label Flipping and Feature Manipulation

**Input:** Training set in each node with risk level in each point $D'_k = \{(x_i, y_i, r_i)\}_{i=1}^n (k = 1,2 ..., K)$, XAI explainer $E$ based on permutation importance, attacking budget $\alpha$, attacking the flag $flag_k = \{0,1\}(k = 1,2 ..., K)$

1: $j \leftarrow 0$
2: $D^* \leftarrow D'$
3: **for** k = 1 to K **do**
4:     **if** $flag_k = 1 (k = 1,2 ..., K)$ **then**
5:         **while** $j < \alpha$ **do**
6:             Extract one data point $(x_i^*, y_i^*, r_i^*)$
7:             **if** $y_i^* = 4$ **then**
8:                 $y_i^* \leftarrow 1$
9:             **else** $y_i^* \leftarrow y_i^* + 1$
10:             **end if**
11:             Extract the most and least important features $f_{max}, f_{min} = E(x_i^*, y_i^*, r_i^*)$
12:             Exchange the most and least important features $E(x_i^*, y_i^*, r_i^*)(E(x_i^*, y_i^*, r_i^*)|x_i^*(f_{max} \leftrightarrow f_{min}))$
13:             Delete $r_i^*$ from data point $(x_i^*, y_i^*, r_i^*)$
14:             $j \leftarrow j + 1$
15:         **end while**
16:     **end if**
17: **end for**

**Output:** Corrupted training set in each node $D_k^*(k = 1,2 ..., K)$.

**Algorithm 3** Reputation-based Federated Learning Defending Algorithm

**Input:** EEG Data set in each node $D_k (k = 1,2 ..., K)$, initial reputation in each node $r_k (k = 1,2 ..., K)$, learning rate $\eta$, initial global parameter $w^0$, initial local parameter $w_k^0 (k = 1,2 ..., K)$, aggregation round T, aggregation group G

1: **for** t = 1 to T **do**
2:     **for** k = 1 to K in parallel **do**
3:         gradient descent to update local models:
4:         $w_k^{t+1} = w^t - \Delta w_k^t = w^t - \frac{\eta}{D_k}[\xi \nabla g(w^t) + \sum_{i=1}^{D_k} \nabla f(w^t, x_{ki}, y_{ki})]$
5:     **end for**
6:     **for** k = 1 to K **do**
7:         calculate the contribution value $e_k^{t+1}$
8:         $e_k^{t+1} = \frac{E[f(w^t, x_{ki}, y_{ki}) - f(w^{t+1}, x_{ki}, y_{ki})]}{E[f(w^t, x_{ki}, y_{ki})]}$
9:         **if** $e_k^{t+1} > e_{min}$ **then**
10:            Add $w_k^{t+1}$ to aggregation group $G^{t+1}$
11:         **else if** $e_k^{t+1} > 0$ **then**
12:             $r_k = r_k \times \frac{e_{min} - e_k^{t+1}}{e_{min}}$
13:         **else**
14:             $r_k = r_k - r_k \times \frac{e_k^{t+1}}{e_{min}}$
15:         **end if**
16:         **if** $r_k < r_{min}$ **then**
17:            Remove $D_k$
18:         **end if**
19:     **end for**
20:     Aggregate global model as:
$w^{t+1} = \sum_{k \in G^{t+1}} (\mu_k^{t+1} w_k^{t+1}), \mu_k^{t+1} = \frac{r_k}{\sum_{i \in G^{t+1}} r_i}$
21:     Broadcast $w^{t+1}$
22: **end for**

**Output:** The final global model parameter $w^T$.

analysis for emotion recognition or sleep stage classification. Feature permutation importance can help researchers understand the workings of their models and enhance their interpretability. After deploying the feature permutation importance method, the rational attack could gain information about the risk level of different EEG signal features and could launch the feature modification attack on the EEG signal points with the highest risk level revealed from Algorithm 1

After attacking stages 1 and 2 stated above, the SVM-XAI attacker has successfully corrupted the EEG signal data set. And from the defenders' side, a reputation-based federated learning system would be established to defend against the attacker.

## IV. THE PROPOSED REPUTATION-BASED FEDERATED LEARNING FRAMEWORK

This section introduces the basics of federated learning followed by the proposed reputation mechanism. The general process of the proposed defending system is shown in Algorithm 3. The management of reputation involves two important tasks, calculating local nodes' reputation by comparing the similarities between global and local nodes' weight, and storing the reputation securely in a decentralized way. A closer view of the proposed reputation-based federated learning defending framework in Algorithm 3 is shown in Fig. 3.

### A. Preliminaries of Federated Learning

Federated learning is a machine learning method that allows mobile devices to work together in training a common model without sharing their private data with a central server. It is a distributed and privacy-preserving technique with a lot of potential [11]. Federated learning can be categorized into different types based on various criteria. One common categorization is based on the type of data used in training: Horizontal Federated Learning, Vertical Federated Learning, and Federated Transfer Learning. In horizontal federated learning, each client has the same features but different examples of data. The goal is to train a model that can generalize across all clients while keeping the data decentralized. In vertical federated learning, different clients have different features but the same set of examples. The goal is to train a model that can learn from all features while keeping the data decentralized. In federated transfer learning, each client has different features and examples of data. The goal is to train a model that can generalize across clients with different distributions of data. In this work, although all training data is collected as EEG signal data, there are major differences among the features and sets of examples of the collected EEG signal datasets. Therefore, the federated learning methods utilized in this work would be categorized into federated transfer learning.

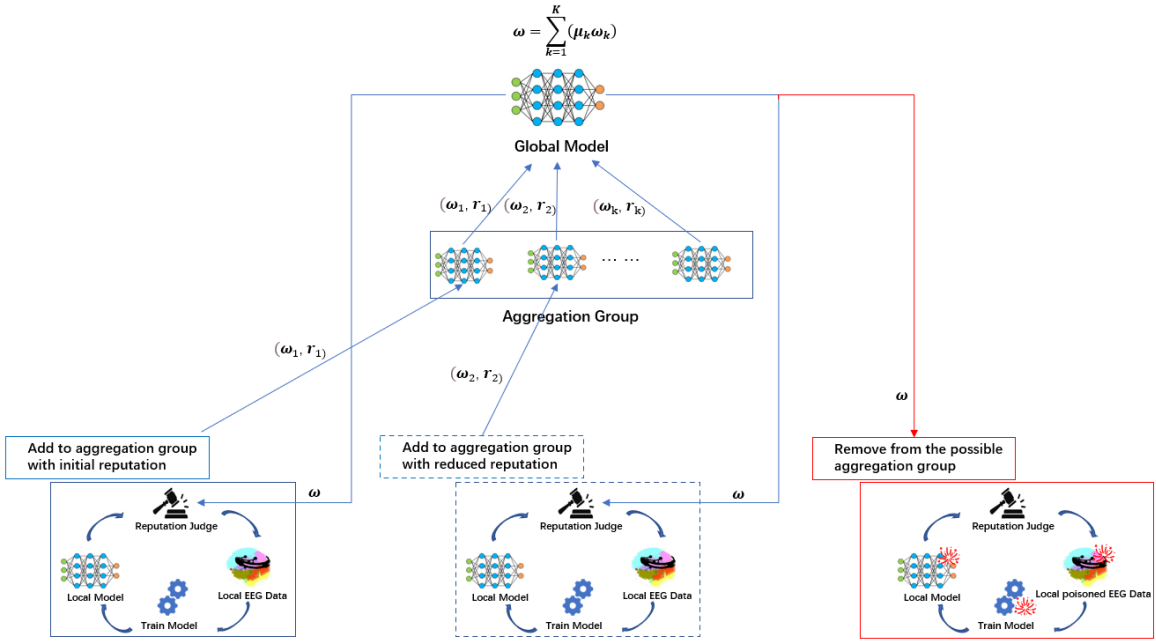

Fig. 3. A closer view of the proposed reputation-based federated learning defending framework in Algorithm 3.

In Federated Learning, the global model is represented as $\Theta$, the total size of data samples is represented as $S = \sum_{k=1}^{K} s_k$ where $K$ is the number of local nodes, loss function of sample data $i$ is represented as $f_i(\Theta)$, and the goal is to optimize a global loss function $\ell(\Theta)$ shown in the below equation [33]:

$$\min_G \ell(\Theta) = \sum_{k=1}^{K} \frac{s_k}{S} \ell_n(\Theta), \text{ where } \ell_n(\Theta) = \frac{1}{s_k} \sum_{i \in s_n} f_i(\Theta). \quad (1)$$

At a training aggregation iteration $t$, $\Delta w_k^t$ represents the average gradient weight of local node $k$ on its dataset. Giving the learning rate $\eta$, the local model update on node $k$ could be expressed by:

$$\Theta_k^{t+1} = \Theta^t - \eta \Delta w_k^t \quad (2)$$

After the updates on the local nodes, the shared global node could be updated by a weighted aggregation of all the local nodes and move to the next iteration, this process could be denoted by:

$$\Theta^{t+1} = \sum_{k=1}^{K} \frac{s_k}{S} \Theta_k^{t+1} \quad (3)$$

The general process flow of a federated learning algorithm has been presented. Nevertheless, the algorithm proposed in this research, as depicted in Algorithm 3, incorporates certain modifications to address the unique features of EEG signal data and reputation mechanism.

*B. Reputation-Based Federated Learning Working Scheme on EEG Signal Data*

To address the challenges of the data poisoning threats proposed by the SVM-XAI attacker in Algorithm 2, this work designs a reputation-based federated learning working scheme. Managing reputation in a federated learning system involves three key tasks: first, calculating the reputation of local nodes by comparing their weights to those of the global node, and second, securely storing the reputation information in a decentralized manner, and then update the global model based on the stored reputation of local nodes. More details about the proposed reputation-based defending framework shown in Algorithm 3 are given as follows.

1) The first step involves the broadcast of federated learning tasks and the initial global model. This process starts with the task publisher broadcasting the task along with specific resource requirements such as data types, data sizes, accuracy, time range, and CPU cycles to mobile devices. Local nodes that satisfy the requirements can become model training worker candidates for the federated learning tasks and respond to the task publisher with their resource information.

2) The second step involves the calculation of the local nodes' reputation in the proposed approach. The global node

calculates the reputation values of these local nodes using the contribution value $e_k^{t+1}$ defined in equation 4. If the contribution value $e_k^{t+1}$ does not exceed the minimum required contribution value $e_{min}$, the reputation value of the node $k$ would be recalculated as equation 5.

$$e_k^{t+1} = \frac{E[f(w^t, x_{ki}, y_{ki}) - f(w^{t+1}, x_{ki}, y_{ki})]}{E[f(w^t, x_{ki}, y_{ki})]} \quad (4)$$

$$r_k = r_k \times \frac{e_{min} - e_k^{t+1}}{e_{min}} \quad (5)$$

3) Once the reputation calculation is completed, the local nodes with a reputation score exceeding a certain threshold can be chosen to add to the aggregation group $G$ with an initial reputation value $r_k$ in this aggregation iteration. On the other hand, the reputation of local nodes with contribution value $e_k^{t+1}$ does not exceed the minimum required contribution value $e_{min}$ would be decreased by equation 5. Once the reputation of the local nodes is lower than the minimum reputation value $r_{min}$, the local nodes with low reputations would be removed from the possible aggregation group nodes.

4) Once the aggregation group selection process is complete, the federated learning aggregation tasks could commence, and the models could be trained using various optimization algorithms such as SGD (Stochastic Gradient Descent). In particular, an initial SGD model is randomly selected from predefined parameter ranges to serve as the shared global model at the start of the training process. And the global model would be aggregated by the reputation weights of the local nodes in the aggregation group described in equation 6.

$$w^{t+1} = \sum_{k \in G^{t+1}} \left( \mu_k^{t+1} w_k^{t+1} \right), \mu_k^{t+1} = \frac{r_k}{\sum_{i \in G^{t+1}} r_i} \quad (6)$$

And the latest global node model would be broadcasted to all possible local nodes, then the next iteration of reputation computing and node selection begins from Step 1.

## V. THE EXPERIMENTAL RESULTS

This section describes the experimental simulation results of the proposed SVM-XAI attacker and the presented reputation-based federated learning defender on some state-of-the-art EEG signal datasets and self-collected EEG signal dataset, KU EEG Brainwave Dataset [34], in terms of emotion recognition based on EEG signal data. The details of the utilized EEG signal datasets are depicted in the following table II.

TABLE II. DETAILS OF UTILIZED EEG SIGNAL DATASETS

| Dataset Name | Subjects | Emotion Labels | Channel numbers | Sampling Rate |
|---|---|---|---|---|
| DEAP [35] | 32 | 40 | 32 | 512 Hz |
| SEED [36] | 15 | 3 | 64 | 256 Hz |
| MAHNOB-HCI [37] | 27 | 4 | 32 | 128 Hz |
| DREAMER [38] | 23 | 7 | 32 | 128 Hz |
| KU EEG Brainwave [34] | 24 | 4 | 25 | 128Hz |

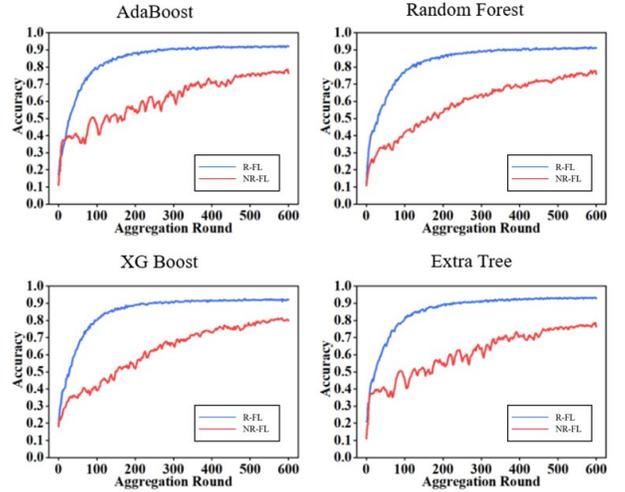

Fig. 4. Rising accuracy curves of different classifiers with Reputation-based Federated Learning and Non-Reputation-based Federated Learning.

As shown in Fig. 3, the SVM-XAI attacker deploys a 20% training data poisoning rate and 4 different machine learning methods, including Random Forest, AdaBoost, XGboost, and Extra Tree are utilized to evaluate the efficiency of the proposed reputation-based federated learning defender. The experimental results show that while the federated learning approach without the reputation-based mechanism could be threatened by the SVM-XAI attacker, the proposed reputation-based federated learning system could mitigate the consequence of the SVM-XAI attacker successfully.

## VI. CONCLUSIONS AND FUTURE WORK

In conclusion, this study demonstrates the effectiveness of the reputation-based defense framework in mitigating security threats in EEG signal classification within the context of Federated Learning. The framework utilizes the collaborative model training approach of Federated Learning to protect privacy by leveraging localized data from various distributed sources. It also introduces a reputation-based mechanism to identify compromised participants and counter the impact of data poisoning attacks. Future research endeavors in this field may involve enhancing the proposed framework by exploring alternative reputation-based mechanisms and assessing their efficacy in addressing security threats. Furthermore, there is scope for investigating the application of Federated Learning in domains beyond EEG signal classification, thus broadening its potential impact and utility.

ACKNOWLEDGMENT

The Authors gratefully appreciate the support from the Khalifa University and Technology Innovation Institute (TII) under Grant PALM 8434000394.

# Authors' background

| Name | Prefix | Research Field | Email | Personal website |
|------|--------|----------------|-------|------------------|
| Zhibo Zhang | Master Student | Federated Learning, Cyber Physical System | 100060990@ku.ac.ae | [Zhibo Zhang - Homepage (qiuyuezhibo.github.io)](https://qiuyuezhibo.github.io) |
| Pengfei Li | Master Student | Computer Vision | 100062567@ku.ac.ae | |
| Ahmed Y. Al Hammadi | PhD Candidate | Cyber Security | ahmed.yalhammadi@ku.ac.ae | |
| Fusen Guo | Master Student | Deep Learning, Smart Grid | dobbysen430@gmail.com | |
| Ernesto Damiani | Full Professor | Intelligent systems | ernesto.damiani@ku.ac.ae | |
| Chan Yeob Yeun | Associate Professor | Cyber Security | chan.yeun@ku.ac.ae | |

**Note:**

[1] **This form helps us to understand your paper better; the form itself will not be published.**

[2] *Prefix*: can be chosen from Master Student, PhD Candidate, Assistant Professor, Lecturer, Senior Lecturer, Associate Professor, Full Professor